\documentclass[aps,prb,twocolumn,letterpaper]{revtex4}

\usepackage{amsmath,amssymb,amsfonts,amsthm}
\usepackage{graphicx,bm}

\begin{document}

\title{High temperature memory in (Pb/La)(Zr/Ti)O$_{3}$ as intrinsic of the
relaxor state rather than due to defect relaxation}

\date{}

\author{F. Cordero,$^1$ F. Craciun,$^1$ A. Franco$^1$ and C. Galassi$^2$}
\affiliation{$^1$ CNR-ISC, Istituto dei Sistemi Complessi, Area
della Ricerca di Roma - Tor Vergata,\\
Via del Fosso del Cavaliere 100, I-00133 Roma, Italy}
\affiliation{$^{2}$ CNR-ISTEC, Istituto di Scienza e Tecnologia della Ceramica,\\
Via Granarolo 64, I-48018 Faenza, Italy}

\begin{abstract}
It has been recently shown that the memory of multiple aging stages,
a phenomenon considered possible only below the glass transition of
some glassy systems, appears also above that temperature range in
the relaxor ferroelectric (Pb/La)(Zr/Ti)O$_{3}$ (PLZT). Doubts exist
whether memory at such high temperature is intrinsic of the glassy
relaxor state or is rather due to migration of mobile defects. It is
shown that the memory in the electric susceptibility and elastic
compliance of PLZT 9/65/35 is not enhanced but depressed by mobile
defects like O vacancies, H defects and mobile charges resulting
from their ionization. In addition, memory is drastically reduced at
La contents slightly below the relaxor region of the phase diagram,
unless aging is protracted for long times (months at room
temperature). This is considered as evidence that in the non relaxor
case memory is indeed due to slow migration of defects, while in the
La rich case it is intrinsic of the relaxor state, even above the
temperature of the susceptibility maximum.
\end{abstract}


\maketitle

\section{Introduction}

Relaxor ferroelectrics are under many aspects the electric analogue of spin
glasses;\cite{Sam03,Kle06} they are closely related to ferroelectrics, but
with enough chemical disorder and/or frustrated interaction to transform the
ferroelectric transition into a gradual freezing of the polar degrees of
freedom, easily recognizable from the broad dielectric susceptibility
maximum at a frequency dependent temperature $T_{m}\left( \omega \right) $.
The characteristic relaxation time $\tau $ deduced from the relationship $%
\omega \tau \sim 1$ at the maximum diverges at a finite temperature $T_{f}$
approximately following the Vogel-Fulcher law $\tau =\tau _{0}\exp \left[
E/\left( T-T_{f}\right) \right] $, or a power law, $\tau =\tau _{0}\left(
\frac{T-T_{f}}{T_{f}}\right) ^{z\nu }$, and $T_{f}$ can be considered as an
indication of the freezing temperature or glass transition. In addition,
these materials exhibit aging, rejuvenation and memory; such phenomena,
usually observed and studied below $T_{f}$, appear in the electric
susceptibility $\chi \left( \omega ,T\right) $ but also elastic compliance $%
s\left( \omega ,T\right) $ as a slow isothermal decrease (aging); on further
cooling, the susceptibilities increase again joining the reference curve $%
\chi _{\text{ref}}\left( \omega ,T\right) $ measured during continuous
cooling (rejuvenation); during reheating, the hole in the $\chi \left(
\omega ,T\right) $ curve created by aging may be partially retraced
(memory). Aging in relaxor ferroelectrics has been found as early as 1980,%
\cite{BCS80} but only recently the non-equilibrium phenomena have started to
be studied in this class of materials. Aging is a rather frequent phenomenon
with various possible causes, but rejuvenation and especially memory are
considered as peculiar of the frozen spin glass state,\cite{NMN00} and
theoretical models for them rely on a proliferation of hierarchically
organized metastable states below a glass transition temperature,\cite{VHO96}
or the vanishing of the barriers between these metastable states above a
temperature $T_{f}$.\cite{BDH01} Memory is therefore expected to vanish
above $T_{f}$, but it has been observed also above that temperature in few
systems: in the magnetic perovskite Y$_{0.7}$Ca$_{0.3}$MnO$_{3}$, which
presents an unusual state of short-range ferromagnetic order before freezing
into a canonical spin glass state;\cite{MNN01} in the relaxor ferroelectric
(Pb/La)(Zr/Ti)O$_{3}$ (PLZT)\cite{CCW01} and to some extent in Pb(Mg$_{1/3}$%
Nb$_{2/3}$)$_{1-x}$Ti$_{x}$O$_{3}$ (PMN-PT)\cite{KB02b} but not in PMN.\cite%
{CCW01} A detailed study\cite{CCW01} showed that the memory recovery of
aging stages at a temperature $T_{a}$ becomes anomalously high when $T_{a}$
approaches $T_{f}$ from below and remains high above $T_{f}$ in PLZT,
instead of vanishing as in PMN or in spin glasses; for this reason,\emph{\ }%
it was proposed that high-temperature aging below and above $T_{f}$ in PLZT
are of different nature, the latter possibly being related to mobile defects
like oxygen vacancies.\ Indeed, such defects are held responsible for the
closely related phenomena of fatigue and domain wall clamping in
ferroelectrics,\cite{RA93} which also involve a slow decrease of the
dielectric susceptibility. On the other hand, we showed\cite{115} that PLZT
has even memory of \textit{multiple} aging stages above $T_{f}$. Memory of
multiple aging stages has never been observed -- and probably not looked for
-- outside a canonical spin glass\cite{JNV00,YLB01} and possibly
superparamagnetic\cite{SSG03b,ZGX05,SJT05} state; therefore, its presence is
already an indication that in PLZT some mechanism of hierarchically
organized correlations intrinsic of the relaxor state is involved also above
the relaxor transition, rather than marginally mobile defects.\cite{115}
Still, in the absence of a quantitative model for the contribution of mobile
defects on aging and memory, the possibility cannot be excluded that defects
are the cause of such non equilibrium phenomena above $T_{f}$, also because
O\ defects up to few tenths of molar percent may arise from the compensation
of unavoidable deviations from the ideal cation stoichiometry.

It is our purpose here to study the influence of O\ vacancies and H
impurities on aging and memory in the region near and above $T_{f}$ in PLZT,
since these are the only defects whose mobility might produce relaxation
effects with characteristic times of the order of hours or days at such
temperatures (300-400~K). In fact, O is by far more mobile than the cations,
with an activation energy for diffusion of $\sim 1$~eV in titanate
perovskites,\cite{PKR99,121} and the only possible O defect is a vacancy,
since there would not be room enough for interstitial O in the closely
packed perovskite structure. Hydrogen may enter as (OH)$^{-}$ ion
substituting an O vacancy and its diffusion barrier is even lower.\cite{ND95}
Regarding the off-centre ionic displacements, they are fully responsible for
the dielectric and relaxor properties of these materials and cannot be
considered as extrinsic defects affecting memory. Finally, mobile charges
from non compensated or ionized defects might be considered responsible for
anomalous non-equilibrium phenomena, but they also produce thermally
activated relaxation processes in the dielectric spectra at high
temperature, that are not observed in PLZT.

\section{Experimental\label{sec Experimental}}

We tested two compositions of (Pb$_{1-x}$La$_{x}$)(Zr$_{y}$Ti$_{1-y}$)$%
_{1-x/4}$O$_{3}$: one, with $x=0.09$, $y=0.65$ (PLZT 9/65/35) and charge
compensating vacancies in the Zr/Ti sublattice, is well within the relaxor
state below $T_{f}<340$~K; such a composition has been extensively
investigated in the literature,\cite{LG77,KBP00,CCW01,115} and the sample
was the same used in previous work where we observed multiple memory at
relatively high temperatures.\cite{115} The other composition was (Pb$%
_{0.93} $La$_{0.07}$)(Zr$_{0.6}$Ti$_{0.4}$)$_{0.96}$Nb$_{0.01}$O$_{3-\delta
} $ (PLZTN 7/60/40), with $x=0.07$, $y=0.60$, and 1\% substitutional Nb and
a slight excess of charge compensating vacancies in the Zr/Ti sublattice;
the cation charge is 5.96 per formula unit, requiring 0.02 O\ vacancies for
the charge neutrality condition. According to the PLZT phase diagram,\cite%
{LG77} PLZTN 7/60/40 is at the border between ferroelectric and relaxor
state below $T_{\mathrm{C}}\simeq 460$~K.

The ceramic samples have been prepared by the mixed-oxide method; the
starting oxide powders were calcined at 850~$^{\mathrm{o}}$C for 4 hours,
pressed into bars and sintered at 1250~${^{\mathrm{o}}}$C for 2~h, packed
with PbZrO$_{3}$ + 5wt\% excess ZrO$_{2}$ in order to maintain a constant
PbO activity at the sintering temperature. The absence of impurity phases
was checked by powder X-ray diffraction. The ingots were cut into thin bars
approximately $45\times 4\times 0.5$~mm$^{3}$. The electrodes for the
anelastic and dielectric spectroscopy measurements were applied with silver
paint and the samples were annealed in air at 700~$^{\mathrm{o}}$C for
avoiding any effects from the possibly damaged surfaces after cutting. The
introduction of O vacancies into PLZT 9/65/35 was first attempted by heating
up to 660~$^{\mathrm{o}}$C in vacuum down to $2\times 10^{-8}$~mbar,
obtaining an irregularly darker sample surface; a more drastic reducing
treatment was therefore made in 800~mbar of H$_{2}$ at 690~$^{\mathrm{o}}$C
for 90~min. The latter treatment changed the sample color into very dark and
uniform gray; assuming that the only effect of the treatment was O\ loss
(without H uptake and no effect on the Ag electrodes), the mass reduction
corresponded to an introduction of $3.3\times 10^{-3}$ O vacancies per
formula unit and, judging from the results of similar treatments on SrTiO$%
_{3}$,\cite{121} it is likely that some H entered the sample.

The dielectric susceptibility $\chi =\chi ^{\prime }-i\chi ^{\prime \prime }$
was measured with a HP 4194 A impedance bridge with a four wire probe and a
signal level of 0.5 V/mm, between 200 Hz and 1 MHz. The measurements were
made on heating/cooling at $\pm 1.5$~K/min between 520 to 240~K in a Delta
climatic chamber. The chamber was enclosed in a sealed box that was flushed
with dry nitrogen in order to prevent the condensation of water on the
sample from air during cooling; otherwise, during subsequent heating the
condensed water would shortcut the electrodes giving rise to anomalies in
the spectra.

The dynamic Young's modulus $E\left( \omega ,T\right) =E^{\prime
}+iE^{\prime \prime }$ or its reciprocal, the elastic compliance $%
s=s^{\prime }-is^{\prime \prime }=E^{-1}$, was measured by electrostatically
exciting the flexural modes of the bars suspended in vacuum on thin
thermocouple wires in correspondence with the nodal lines; the 1st, 3rd and
5th modes could be measured, whose frequencies are in the ratios $1:5.4:13.3$%
, the fundamental frequencies of the PLZT 9/65/35 and PLZTN 7/60/40 samples
being $\omega /2\pi \simeq 1$ and 0.7~kHz respectively. The real part of the
Young's modulus is related to the resonance frequency through $\omega
_{n}=\alpha _{n}\sqrt{E^{\prime }/\rho }$, where $\alpha _{n}$ is a
geometrical factor of the $n$-th vibration mode and $\rho $ the mass
density. The latter varies with temperature much less than $E^{\prime }$, so
that $\omega ^{2}\left( T\right) /\omega ^{2}\left( T_{0}\right) \simeq $ $%
E\left( T\right) /E\left( T_{0}\right) =$ $s\left( T_{0}\right) /s\left(
T\right) $. The elastic energy loss coefficient, or the reciprocal of the
mechanical quality factor,\cite{NB72} is $Q^{-1}\left( \omega ,T\right) =$ $%
E^{\prime \prime }/E^{\prime }=$ $s^{\prime \prime }/s^{\prime }$; the $%
Q^{-1}$ was measured from the decay of the free oscillations or from the
width of the resonance peak. The elastic compliance $s$ is the mechanical
analogue of the dielectric susceptibility $\chi $, with $Q^{-1}$
corresponding to $\tan \delta $. The absolute value of $s$ is difficult to
be evaluated, due to sample porosity, slightly irregular shape and to the
contribution from the Ag electrodes; therefore the $s^{\prime }$ curves will
be normalized with respect to the value $s_{0}$ extrapolated to very high
temperature. The heating and cooling rates during the anelastic experiments
were of $0.8-1$~K/min. All the anelastic and dielectric measurements started
from above 500~K, in order to erase any previous history.

\section{Results}

\subsection{The relaxor PLZT 9/65/35}

\subsubsection{Dielectric measurements}

\begin{figure}[hbpt]
\centerline{\includegraphics[width=\columnwidth]{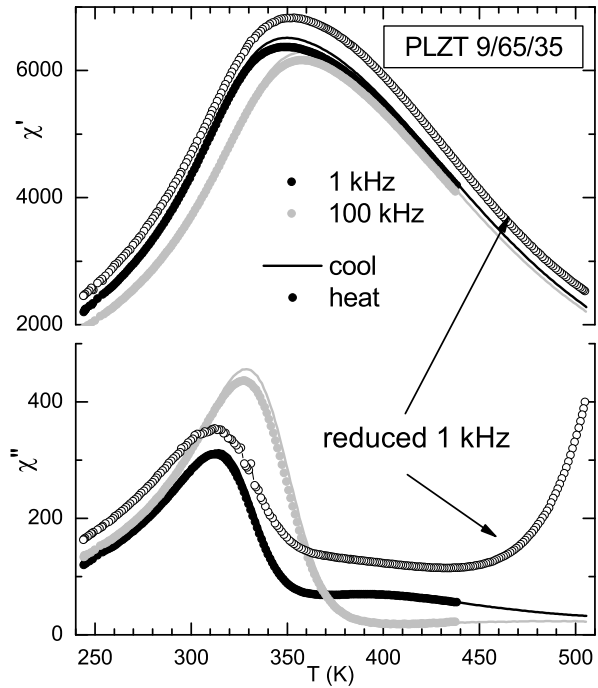}}
\caption{Real and imaginary dielectric susceptibility of PLZT
9/65/35 measured during cooling and heating at $\pm 1.5$~K/min
(reference curves). Black and gray refer to 1 and 100~kHz
respectively. The empty symbols are the cooling curves after
reducing the sample in H$_{2}$ atmosphere.}
\label{fig diel96535}
\end{figure}

The dielectric constant $\chi ^{\prime }$ and losses $\tan \delta =\chi
^{\prime \prime }/\chi ^{\prime }$ of PLZT 9/65/35 are shown in Fig. \ref%
{fig diel96535}, measured both on cooling down to 240~K at $-1.5$~K/min
(continuous lines)\ and on subsequent heating at 1.5~K/min (filled symbols).
Only two frequencies, 1 and 100~kHz, are shown, but the measurements were
extended between 200~Hz and 1~MHz, and gave the usual result found in the
literature.\cite{DXL96,KBP00,115} Notice the merging into a single curve of
the $\chi ^{\prime }$ curves at high temperature and of the $\chi ^{\prime
\prime }$ ones at low temperature (see also Refs. \onlinecite{115,116} on
the same and similar samples), characteristic of the spin glass
susceptibilities. As usual for glassy systems, it is possible to define a
characteristic relaxation time $\tau $ from the condition $\omega \tau
\left( T_{m}\right) =1$ at the maximum of $\chi ^{\prime }$, and to fit it
with a Vogel-Fulcher law, $\tau =$ $\omega =$ $\tau _{0}\exp \left[ E/\left(
T_{m}-T_{f}\right) \right] $ or a power law,\emph{\ }$\tau \sim \left( \frac{%
T-T_{f}}{T_{f}}\right) ^{-z\nu }$, where the temperature $T_{f}$ at which
the characteristic time diverges indicates the freezing temperature or the
glass transition. This analysis with the present samples yields\cite{115} $%
T_{f}=$ 320~K for the Vogel-Fulcher and 337~K\emph{\ }and $z\nu =6.8$ for
the scaling approach, without the possibility of clearly distinguishing
which one is the better description; it is however clear that the freezing
temperature for the polar fluctuations is lower than 340~K. Also shown in
Fig. \ref{fig diel96535} with empty symbols is $\chi $ measured at 1~kHz on
cooling, after the reduction treatment in H$_{2}$ atmosphere. The peak in
the real part is slightly shifted to lower temperature and enhanced in
intensity, while the losses acquire an intense thermally activated
contribution at high temperature, attributable to the relaxation of mobile
charges introduced by the O and H defects.

A weak relaxation in $\chi ^{\prime \prime }$, with maximum around 400~K at
1~kHz, appears also in the oxygenated state; this was originally absent, as
shown in Ref. \onlinecite{115}, and is probably due to the fact that the
sample has been repeatedly subjected to reducing and oxygenation treatments,
from which it did not recover completely. This fact does not weaken the
present conclusions, since the memory of even multiple aging stages is found
also in the pristine state\cite{115}, and the "oxygenated" and "reduced"
state considered here certainly differ considerably in the amount of
defects, as demonstrated by the curves in Fig. \ref{fig diel96535} and the
changes in sample color.

\subsubsection{Dielectric aging, rejuvenation and memory}

\begin{figure}[hbpt]
\centerline{\includegraphics[width=\columnwidth]{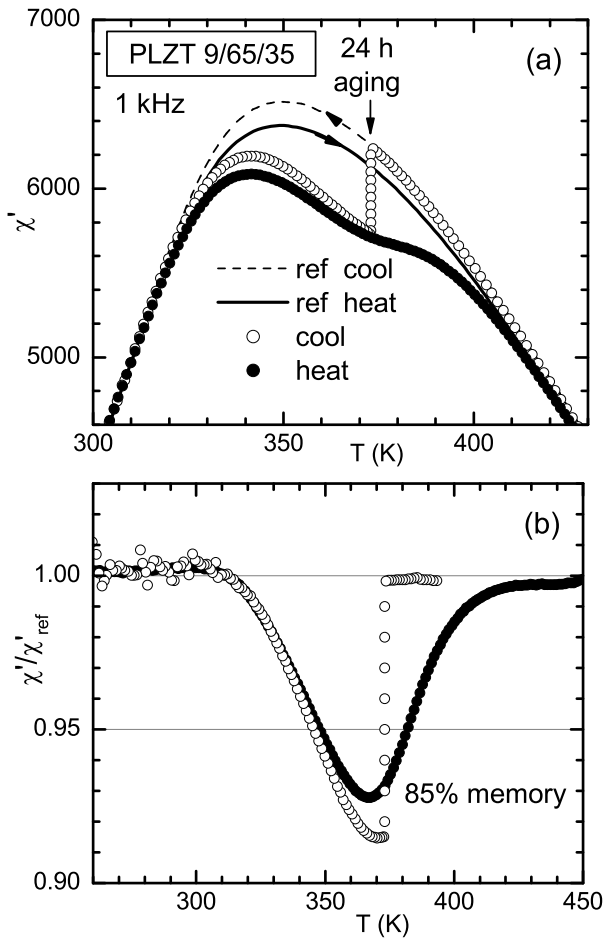}}
\caption{Aging, rejuvenation and memory of PLZT 9/65/35. (a) raw
data, including the reference curves measured during continuous
cooling and heating; (b) data normalized with respect to the
corresponding reference curves.}
\label{fig mem96535}
\end{figure}

Figure \ref{fig mem96535} presents the effect of aging 24~h at 373~K, well
above the freezing temperature or glass transition, $T_{f}<340$~K. The
quality of the data is better in $\chi ^{\prime }$ rather than in $\tan
\delta $, especially after the reduction treatment, and therefore we will
consider the $\chi ^{\prime }\left( \omega ,T\right) $ curves measured at $%
\omega /2\pi =1$~kHz, but very similar results are obtained at the other
frequencies.

After dividing by the cooling and heating reference curves (Fig. \ref{fig
diel96535} and continuous lines in Fig. \ref{fig mem96535}a), one obtains
the left half of a negative peak from the cooling curve after aging and a
slightly smaller peak from the heating curve, where the fraction of memory
recovery can be defined as the ratio between the intensities of the two
peaks. The case of the well oxygenated PLZT\ 9/65/35 is rather clear-cut,
with the $\chi ^{\prime }\left( \omega ,T\right) $ curves overlapping almost
perfectly (better than 0.3\%) onto the reference curves below the aging
temperature (perfect rejuvenation) and above it. We operatively define the
memory recovery in a manner that works also for the less regular curves
obtained after reduction, as the ratio
\begin{equation}
r=\frac{\chi ^{\prime }\left( T_{a}\right) /\chi _{\text{ref c}}^{\prime
}\left( T_{a}\right) }{\chi ^{\prime }\left( T_{\min }\right) /\chi _{\text{%
ref h}}^{\prime }\left( T_{\min }\right) }\;,
\end{equation}%
where $\chi _{\text{ref c}}^{\prime }$ and $\chi _{\text{ref h}}^{\prime }$
are the reference curves on cooling and heating, $\chi ^{\prime }\left(
T_{a}\right) $ is the dielectric susceptibility at the end of aging at $%
T_{a}=373$~K and $T_{\min }$ is the temperature of the minimum of $\chi
^{\prime }\left( T\right) $ on heating, slightly smaller than $T_{a}$. This
is the most straightforward definition of memory recovery, taking into
account the fact that the curves measured on heating are slightly lower than
those measured on cooling. Applying this criterion to all the measured
frequencies between 200~Hz and 1~MHz, one gets a memory recovery of $%
84.6\%\pm 0.5\%$ for the dielectric susceptibility.

The experiment was repeated on a sample reduced in H$_{2}$, as described in
Sect. \ref{sec Experimental}, and therefore with a considerably higher
amount of O\ vacancies and mobile charges, now appearing in the dielectric
losses at high temperature (Fig. \ref{fig diel96535}).

\begin{figure}[hbpt]
\centerline{\includegraphics[width=\columnwidth]{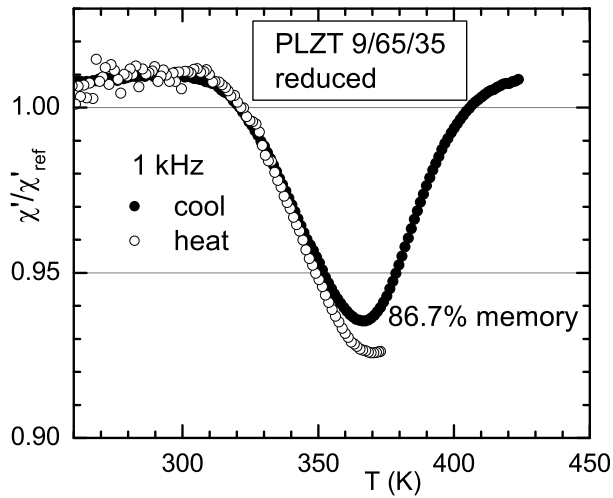}}
\caption{Aging, rejuvenation and memory of a PLZT 9/65/35 sample
after reduction in H$_{2}$ atmosphere, to be compared with Fig.
\protect\ref{fig mem96535}b. The $\protect\chi _{\text{ref}}^{\prime
}$ reference curve on cooling is shown in Fig. \protect\ref{fig
diel96535}.}
\label{fig memdielVO}
\end{figure}

In the reduced sample, the rejuvenation curve is $\sim 1\%$ higher than the
reference near the aging hole, and remains high during heating. Applying the
above criterion to all the measured frequencies one gets a memory recovery
of $86.6\%\pm 0.5\%$, only 2\% higher than in the defect free material.
Considering the sensitivity of the $\chi \left( T\right) $ curves to changes
in the temperature rate, especially in the reduced state, we can say that
the introduction of O and H\ related defects affects the aging behavior but
there is no enhancement of the memory recovery within experimental error.

\subsubsection{Anelastic measurements \label{sec anel ref}}

The same type of experiments has been made with the dynamic compliance,
which has already been shown to exhibit closely similar non-equilibrium
effects.\cite{115} There are however some differences due to the fact that
\textit{i)} there is a strong contribution of the lattice to $s^{\prime }$,
besides that related to the polar fluctuations, and therefore the fraction
of $s^{\prime }$ that exhibits aging and memory is smaller than in the
dielectric experiment, by more than three times in the present case; \textit{%
ii)} the dynamic compliance is sensitive to the strain fluctuations, of
quadrupolar nature, rather than directly to the electric dipolar ones, and
the maximum of the elastic losses is shifted to lower temperature;\cite{116}
therefore, it becomes very difficult to reliably measure aging and memory
phenomena in the $Q^{-1}\left( \omega ,T\right) $ curves at $T\geq 350$~K,
where the losses become comparable with the background. For these reasons,
we chose to study the effect of O vacancies in PLZT 9/65/35 at 326~K both in
the $Q^{-1}\left( T\right) $ and $s^{\prime }\left( T\right) $ curves
measured at 1~kHz. This temperature is 50~K lower than that of the
dielectric experiment, but still within the temperature range where Colla
\textit{et al.} found that the memory recovery becomes anomalously high,
instead of vanishing as in PMN.\cite{CCW01}

\begin{figure}[hbpt]
\centerline{\includegraphics[width=\columnwidth]{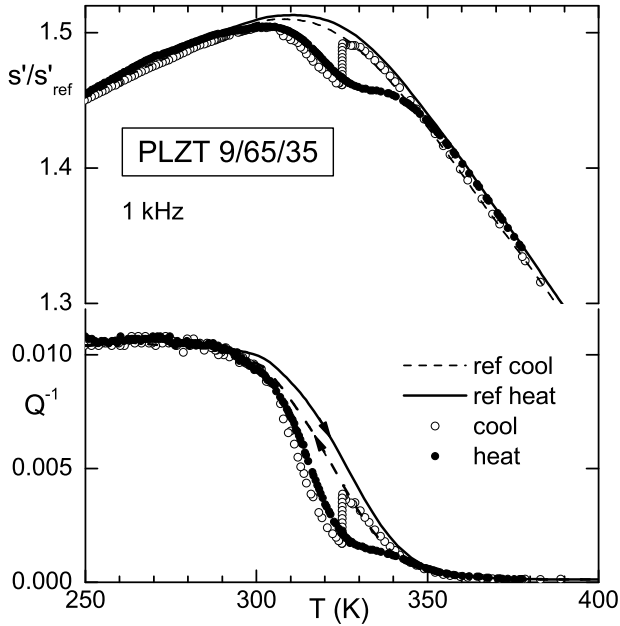}}
\caption{Normalized dynamic compliance $s^{\prime }/s_{0}$ and
elastic energy loss coefficient $Q^{-1}=s^{\prime \prime }/s^{\prime
}$ of PLZT measured at $\simeq 1$~kHz. Broken and continuous lines:
cooling and heating reference curves; open and closed symbols:
cooling with 26~h aging and rejuvenation and heating with memory.}
\label{fig anmem96535}
\end{figure}

\begin{figure}[hbpt]
\centerline{\includegraphics[width=\columnwidth]{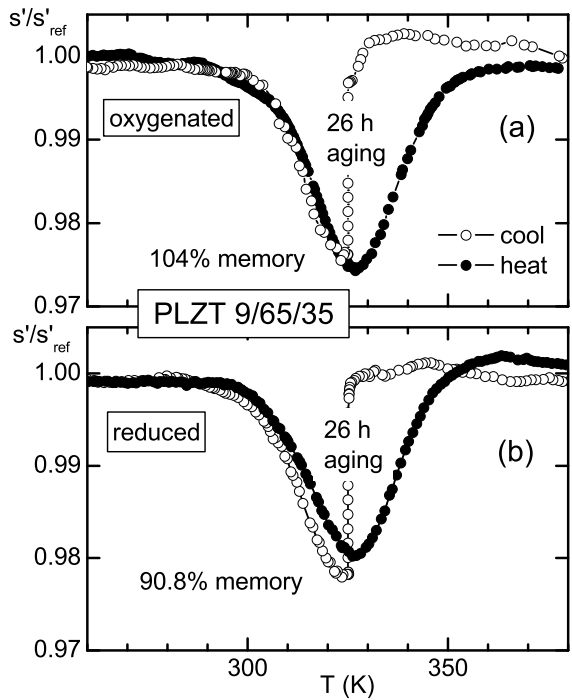}}
\caption{Aging, rejuvenation and memory of PLZTN 7/60/40 after
normalization with the cooling and heating reference curves (see
Fig. \protect\ref{fig annorm96535}).} \label{fig annorm96535}
\end{figure}

Figure \ref{fig anmem96535} shows the raw data of the cooling and heating
reference curves of the normalized dynamic compliance $s^{\prime }/s_{0}$
and elastic energy loss coefficient $Q^{-1}=s^{\prime \prime }/s^{\prime }$
at $\simeq 1$~kHz, together with the effect of aging for 26~h at 325~K,
rejuvenation and memory. The result of normalization with the reference
curves is shown in Fig. \ref{fig annorm96535}(a) for $s^{\prime }$; the $%
Q^{-1}$ curves give exactly the same result but are more noisy. The fact
that the data on cooling before aging differ from 1 is due to the difficulty
of stopping the temperature of the sample suspended in vacuum, but this is
inessential after 26~h of aging. The memory recovery is total, or even
104\%\ by applying the criterion used for the dielectric memory. The result
of the same experiment after reduction in H$_{2}$ atmosphere is shown in
Fig. \ref{fig annorm96535}(b); in this case the memory recovery is
definitely lowered of at least 10\%.

\subsection{The nearly ferroelectric PLZTN 7/60/40}

Figure \ref{fig dielN76040} presents the reference dielectric curves of the
PLZTN 7/60/40 sample, again at 1 and 100~kHz. As expected from the phase
diagram of PLZT, this composition is at the border between the relaxor and
the ferroelectric states:\cite{LG77,DXL96}\ the dispersion in frequency of
the maximum in $\chi ^{\prime }$ is very weak, and the curves measured on
heating (full circles)\ are lower than those measured during cooling
(continuous lines) up to $T_{\text{FE}}=370$~K, indicating the formation of
a partial ferroelectric order, which disappears on heating above $T_{\text{FE%
}}$.

\begin{figure}[hbpt]
\centerline{\includegraphics[width=\columnwidth]{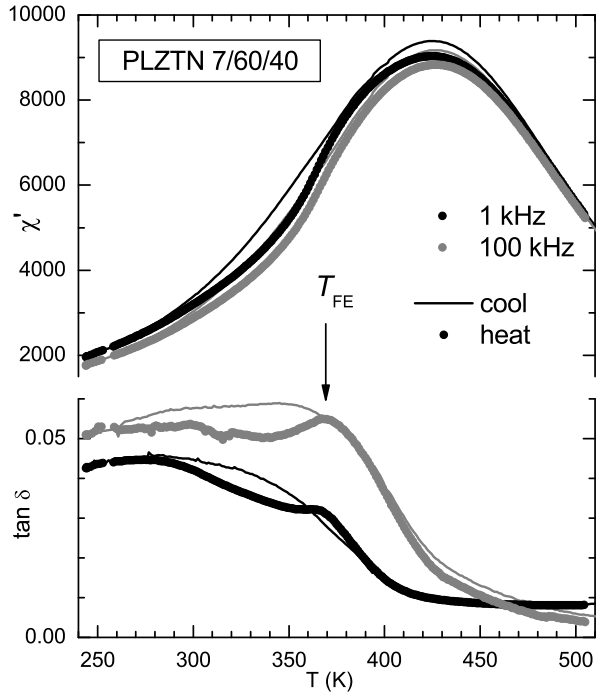}}
\caption{Dielectric susceptibility and losses of PLZTN 7/60/40
measured during cooling and heating at $\pm 1.5$~K/min.} \label{fig
dielN76040}
\end{figure}

\begin{figure}[hbpt]
\centerline{\includegraphics[width=\columnwidth]{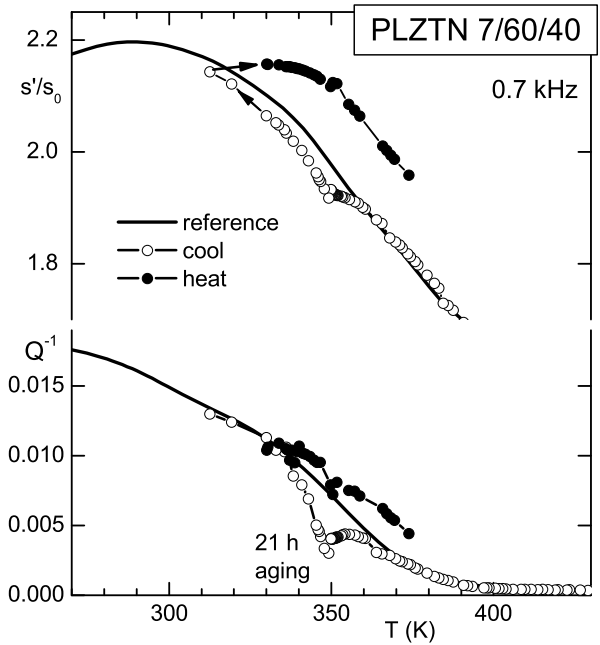}}
\caption{Normalized elastic compliance $%
s^{\prime }$ and elastic energy loss coefficient $Q^{-1}$of PLZTN
7/60/40 measured at 0.7~kHz during continuous cooling at -0.8~K/min
(line); cooling with aging at 350~K (empty circles) and subsequent
heating (filled circles).}
\label{fig shortaging}
\end{figure}

The result of aging 21~h at 350~K is shown in Fig. \ref{fig shortaging} for
the compliance; the initial aging rate is fast compared to the relaxor
phase, as appears from the departure from the reference cooling measurement
already below 360~K, when the cooling rate was slowed down to stop at $%
T_{a}=350$~K. After aging, the cooling was stopped already 35~K below $T_{a}$%
, but this was sufficient to erase almost completely the memory of the aging
stage; the recovery of the hole in the $Q^{-1}$ curve was no more than 30\%,
and that in $s^{\prime }$ was almost unmeasurable, also due to significant
departure from the reference curve. We did not make a systematic study of
the influence of the amplitude of the temperature excursion below $T_{a}$ on
the memory recovery, but it is clear that memory in this partly relaxor and
partly ferroelectric composition is far less clean and complete than in the
pure relaxor phase, where there is practically no effect of the low
temperature excursion, and there is good reproducibility between heating and
cooling curves outside the aging hole.

\begin{figure}[hbpt]
\centerline{\includegraphics[width=\columnwidth]{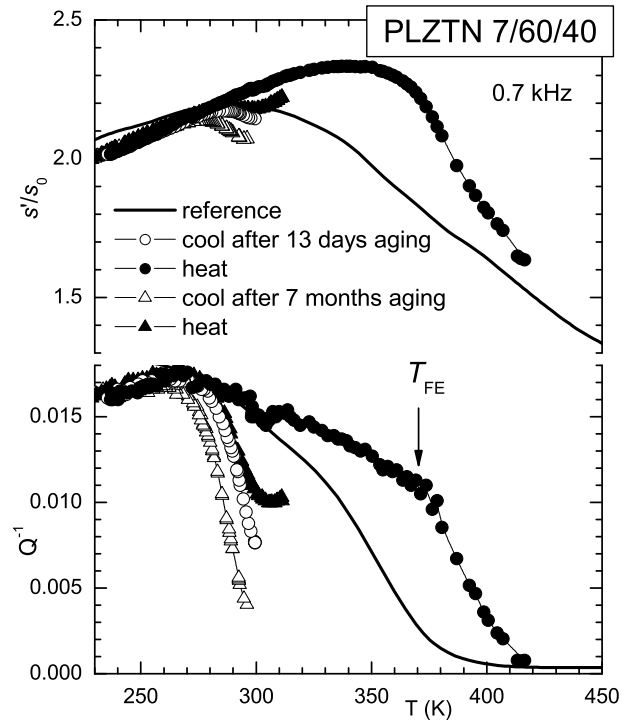}}
\caption{Normalized compliance and elastic energy loss coefficient
of PLZTN\ 7/60/40 after 13 days \ aging at room temperature +
cooling down to 225~K (circles) and 7.5 months aging + cooling to
235~K (triangles). The continuous line is the reference curve on
cooling.} \label{fig longaging}
\end{figure}

This fact can be seen also in Fig. \ref{fig longaging}, where the
effect of 13 days aging at room temperature is almost completely
erased by cooling 70~K below $T_{a}$ (circles, the memory is no more
than 15\%). Note that in this case it is $T_{a}\sim 0.7T_{m}$,
$T_{m}\simeq 430$~K being the temperature of the maximum in
dielectric susceptibility, and at $0.7T_{m}$ relaxor PLZT 9/65/35
and spin-glass systems have full memory recovery. A more substantial
and stable memory is observed after 7.5 months aging at room
temperature (triangles, $\sim 70\%$ recovery). Note also that the
losses on heating remain higher than during cooling, but abruptly
decrease above the same $T_{\text{FE}}$ determined by the dielectric
measurement.

\section{Discussion}

The aim of these experiments is to evaluate the possible contribution of
extrinsic defects to the memory of aging stages found in the relaxor PLZT
9/65/35 also near and above the freezing temperature $T_{f}$. The following
discussion does not rely on a precise definition of $T_{f}$ for PLZT, and we
will also assimilate it to the temperature $T_{m}$ of the maximum of $\chi
^{\prime \prime }$ at some low frequency.\cite{CCW01} The reason why one
might expect some mechanism extrinsic with respect to the relaxor
transition, is that it is generally accepted that aging to some extent and
particularly rejuvenation and memory need some hierarchical organization of
the microscopic states and interactions, which is expected to arise only
below a glass transition.\cite{NMN00,SN00,DVB01,MPP01,SJT05}

An extensive investigation on the memory in relaxors has been conducted by
Colla and coworkers\cite{CCW01} on PLZT 9/65/35, PMN and PMN-PT. These
authors found that PMN behaves like a canonical spin glass, with its memory
of an aging stage at $T_{a}$ vanishing when $T_{a}$ approaches $T_{m}$ from
below. Instead, the memory of PLZT 9/65/35 has an initial decrease with a
minimum around $T_{a}=0.85T_{m}$ and then rises again to 80\%\ at $T_{m}$,
remaining high up to $1.15T_{m}$. The comparison of the memory recovery
versus $T_{a}$ in PMN and PLZT together with the above considerations indeed
suggests that a different mechanism exists above $T_{f}$ in PLZT, which the
authors identify with mobile defects.

To our knowledge, the only defects whose mobility may be high enough to
match with the temperature and characteristic times of these aging
experiments are:\ mobile charges (electrons or holes)\ from ionized defects,
O\ vacancies and hydrogen, generally associated in the form of (OH)$^{-}$
ion in perovskites.\cite{ND95} In the present study, we tested all these
possibilities.

The reference relaxor composition has a rather clear-cut behavior, typical
of spin-glasses:\cite{CCW00,BDH01,115} almost overlapping cooling and
heating reference $\chi \left( \omega ,T\right) $ curves, with no frequency
dependence in the low-temperature $\chi ^{\prime \prime }\left( \omega
,T\right) $ and high-temperature $\chi ^{\prime }\left( \omega ,T\right) $
(Fig. \ref{fig diel96535}), complete rejuvenation on cooling, heating curves
with substantial or complete memory recovery and good matching with the
reference far from $T_{a}$; in addition, one day aging is sufficient to
imprint a complete (Fig. \ref{fig annorm96535}a) or almost complete (Fig. %
\ref{fig mem96535}) memory of the aging stage. The only peculiarity is that
memory exists both below and above $T_{f}$. Now, if the high temperature
memory were almost completely due to defects, it should be considerably
enhanced after reduction in H$_{2}$. In fact, such a treatment introduced at
least $3.3\times 10^{-3}$ O vacancies per formula unit, as deduced from the
mass loss, and most likely also H defects, whose amount we cannot evaluate;
as a result, the sample turned from white-yellow to dark grey and an
extremely intense thermally activated relaxation process appeared in the
high temperature side of $\chi ^{\prime \prime }$, presumably due the motion
of charges from the new ionized defects. The reduced sample therefore
contains a substantially larger amount of all the mobile defects considered
before, with respect to the standard oxygenated state. The effect on the
memory recovery of high temperature aging, however, is negligible in the
dielectric case (2\% increase, Fig. \ref{fig memdielVO}) and consists in a
depression of over 10\% in the anelastic one (Fig. \ref{fig annorm96535});
note that during the anelastic experiment aging has been done at a
temperature 50~K lower than in the dielectric one, and this should be the
main reason for the larger initial memory recovery and larger loss after
reduction.

Regarding the differences between the dielectric and anelastic measurements,
it can be observed that the dynamic compliance $s$ probes quadrupolar strain
fluctuations, while the dielectric susceptibility probes dipolar
polarization fluctuations, and therefore the $\chi $ and $s$ curves may
differ.\cite{116} In principle, there may be relaxation modes appearing in $%
s $ and not in $\chi $ and \textit{vice versa},\textit{\ }but in practice
there is correlation between strain and polarization fluctuations. For
example, in a perfect crystal the inversion of an electric dipole causes no
change in strain (which is centrosymmetric), while rotations of the O
octahedra cause strain but no polarization changes; in a disordered crystal
like PLZT, however, there are no pure inversions of dipoles or pure
rotations of octahedra, and different modes are coupled together.\cite{116}
We do not know whether, between anelastic and dielectric responses, one is
more significant than the other for studying nonequilibrium processes.
Certainly, the dielectric response is more directly related to the dipolar
interactions, which presumably govern the dynamics of relaxor
ferroelectrics; on the other hand, the anelastic response truly probes the
bulk and is insensitive to mobile charges, unlike $\chi $ in reduced or
defective samples.\cite{LBP02b}

In the second part of the experiment, we tested the composition PLZTN
7/60/40, which, although very close to PLZT 9/65/35, presents differences
with respect to the non-equilibrium phenomena: \textit{i)} PLZT 9/65/35
supports a full relaxor state and in the standard oxygenated state has no
detectable extrinsic mobile defects, like O\ vacancies; \textit{ii)} PLZTN
7/60/40 presents both relaxor-like susceptibility maximum and, below $T_{%
\text{FE}}=$ 370~K, weak ferroelectric properties and should have about 2\%\
of O vacancies in order to balance the cation charge (see Experimental).
Therefore, besides having a large amount of defects, PLZTN 7/60/40 starts
developing ferroelectric domains below $T_{\text{FE}}$; if anomalous memory
phenomena may arise from mobile defects interacting with polar nanodomains
and/or more developed ferroelectric domains, then such phenomena should be
particularly well developed in PLZTN 7/60/40. Again, the experiment shows
that this is not case; rather, the non equilibrium processes related to the
relaxor state and to mobile defects have quite different phenomenologies and
time scales. Figure \ref{fig shortaging} shows that the memory of an aging
stage of 21~h at $T_{a}=$ 350~K is drastically lower than in the full
relaxor case: the value of $T_{a}$ is included between those of the
dielectric and anelastic aging experiments in PLZT 9/65/35, so that defects
have the same mobility in all the experiments; still, even though cooling is
extended only to 35~K below $T_{a}$, compared to the 100~K of the previous
cases, the memory recovery is less than 30\%. In addition, when the
formation of ferroelectric domains, although small and disordered, starts,
the behavior on cooling and heating is far less reproducible than in the
relaxor case. Figure \ref{fig longaging} shows the effect of extending
cooling down to 230~K, as in the experiments with PLZT 9/65/35, after longer
aging stages at room temperature: a 13 days aging causes a marked decrease
of both $s^{\prime }$ and $s^{\prime \prime }$, but negligible memory on
heating. This demonstrates that the combination of small or even more
developed ferroelectric domains and mobile defects, although causes aging,
is not able to create the memory effects found in the relaxor state. Rather,
the slow reorientation and migration of O vacancies within the strain and
electric fields of ferroelectric domains may be responsible for the memory
imprinting on the much longer time scale of months (triangles in Fig. \ref%
{fig longaging}). The closely related phenomena of domain wall clamping%
\cite{RA93} and formation of internal fields\cite{Ren04} from mobile defects
adapting themselves to domains are known to occur in ferroelectrics.\cite%
{Dam98}

\subsection{Possible origin for memory above the freezing temperature}

It remains to be explained how is it possible that in PLZT memory of even
multiple aging stages, as also shown in previous work,\cite{115} may occur
above the temperature of the maximum in the susceptibility. The reason for
this behavior may reside in the fact that, unlike spin glasses, the polar
degrees of freedom in relaxor ferroelectrics freeze out of an already
partially correlated state. In fact, relaxor ferroelectrics are
characterized by the Burns temperature $T_{\text{B}}\gg T_{f}$, below which
the formation of polar nanoclusters starts;\cite{Sam03,Kle06} therefore, the
state below $T_{\text{B}}$ is a sort of supeparaelectric one, and the
analogy between relaxors and magnetic systems is probably more appropriate
with strongly correlated superparamagnets rather than with canonical
spin-glasses, where the dynamics above $T_{f}$ is that of individual spins,
although interacting. Indeed, the magnetic perovskite Y$_{0.7}$Ca$_{0.3}$MnO$%
_{3}$, presents a spin glass state preceded by short-range ferromagnetic
order, reminiscent of the polar nanoclusters of relaxors, and also exhibits
memory above the spin glass transition.\cite{MNN01} The authors found this
behavior unusual and were not able to classify it according to any available
theory, but suggested that it may be related to the phase-separated nature
of manganites. In ferroelectric relaxors no electronic phase separation
occurs, but certainly the analogy with Y$_{0.7}$Ca$_{0.3}$MnO$_{3}$ suggests
that highly correlated and hierarchically organized states may arise also
from the interactions among the polar or magnetic clusters, which on further
cooling freeze into the relaxor or spin-glass state. It should be remarked
that, in order to explain memory, the need of hierarchical or at least
strongly correlated dynamics, opposed to simple broad distribution of
relaxation times, is not only indicated by the available theoretical models,
but also by experiments on superparamagnets.

Magnetic nanoparticle systems can be made with a controlled distribution of
particle sizes and with different degrees of dipolar interaction among the
particles. It is therefore possible to obtain slow dynamics in systems
ranging between two limits: \textit{i)} weakly interacting particles with a
broad distribution of individual relaxation times, due to a broad
distribution of their sizes, and \textit{ii)} strongly interacting particles
with collective dynamics. Both limits exhibit various forms of
non-equilibrium dynamics below some blocking temperature $T_{b}$, which from
the phenomenological point of view corresponds to $T_{f}$ in spin-glasses.
Even systems belonging to the first category may exhibit some kind of
memory; memory of multiple aging stages has been found in the dc
magnetization of weakly interacting superparamagnets,\cite{SSG03b,ZGX05}
where the $M\left( T\right) $ curve on heating retraces, although smoothly,
the steps formed during previous field cooling with aging stages at zero
field. It has been remarked, however, that such a behavior can be completely
accounted for in terms of broad distribution of relaxation times, and that
these systems lack memory in the ac susceptibility under zero field.\cite%
{ZGX05} Instead, memory in zero field cooled experiments can be found only
in strongly interacting superparamagnets, also called superspin glasses.\cite%
{SJT05} The analogy between relaxor PLZT and superspin glasses, however, is
not straightforward, since the latter exhibit aging and memory below a
blocking temperature that is the equivalent of $T_{f}$ in spin glasses, and
we are not aware of experiments in such systems where it has been verified
that these phenomena occur also above the temperature of the maxima in the
ac susceptibilities.

A\ successful attempt in modeling aging and memory in the dynamic
susceptibility has been made by Sasaki and Nemoto,\cite{SN00} through the
Multi-layer Random Energy Model; the main hypothesis is that the states of
the system are hierarchically organized into "layers", where the $n$-th
layer contains states that can access the $n+1$-th layer by overcoming
energy barriers $E$ distributed according to $p\left( E\right) \propto \exp
\left( -E/T_{\mathrm{C}}\left( n\right) \right) $, with $T_{\mathrm{C}%
}\left( n+1\right) >T_{\mathrm{C}}\left( n\right) $ and $1\leq n\leq L$. In
this manner it is possible to reproduce by numerical simulation memory
effects in $\chi ^{\prime \prime }\left( t\right) $ during aging stages
below $T_{\mathrm{C}}\left( L\right) \equiv T_{f}$. It would be interesting
to also know the shapes of the $\chi \left( \omega ,T\right) $ curves
resulting from this model, in order to correlate the maximum temperature at
which memory is found with the maxima in $\chi \left( \omega ,T\right) $.
The $\chi \left( \omega ,T\right) $ in the limiting case of a single layer
can be analytically approximated,\cite{LC95} and has been used to describe
the susceptibility of PMN-PT, but does not describe PLZT;\cite{116} in this
case the maxima of $\chi \left( \omega ,T\right) $ are above $T_{\mathrm{C}}$%
.

\section{Conclusions}

The purpose of the present study was to ascertain whether mobile defects,
mainly O\ vacancies, interacting with polar nanoclusters or microdomains
might be the origin of the memory in the susceptibility curves versus
temperature of aging stages also above the relaxor transition; both
dielectric susceptibility and elastic compliance have been tested.

In one experiment, we started from the well-behaved relaxor composition PLZT
9/65/35, presenting from 80\% to full memory recovery of aging stages both
in the dielectric and elastic curves, and introduced all conceivable types
of extrinsic mobile defects:\ O vacancies, hydrogen defects and free charges
from the ionized defects. As a result, the memory was almost unaffected or
depressed, excluding that such defects are the main responsible for the
high-temperature memory.

In another experiment we tested PLZTN 7/60/40, whose composition is very
close to PLZT 9/65/35, but has mixed relaxor and ferroelectric properties
and, due to the cation stoichiometry, should contain about 2~mol\% O
vacancies to ensure charge neutrality. This composition presents almost no
memory imprinting, on the time scale of the previous experiments (1 day),
again demonstrating that the interaction between mobile defects and polar
clusters or ferroelectric microdomains cannot give rise to memory phenomena
as found in relaxors and spin glasses. On the other hand, memory in the
defective system is imprinted on a much longer time scale (months),
suggesting that in this case the migration and reorientation of defects in
the field of well developed ferroelectric domains occurred.

These findings show that the memory phenomena typical of certain canonical
spin glasses and found in some relaxor ferroelectrics are intrinsic of the
relaxor state also above the freezing temperature for the polar
fluctuations. A comparison is made with Y$_{0.7}$Ca$_{0.3}$MnO$_{3}$, where
the spin glass state is preceded by short-range ferromagnetic order,\cite%
{MNN01} and with superparamagnetic systems, where it is possible to tune
both the width of the relaxation times of the relaxing units and their
degree of interaction.


\end{document}